\documentclass[11pt,a4paper]{article}

\usepackage[utf8]{inputenc}
\usepackage[T1]{fontenc}
\usepackage{geometry}
\usepackage{amsmath,amssymb,amsthm}
\usepackage{booktabs}
\usepackage{graphicx}
\usepackage{adjustbox}
\usepackage{hyperref}
\usepackage[auth-lg]{authblk} 
\usepackage[numbers,sort&compress]{natbib} 
\usepackage{color}

\graphicspath{{Fig/}}

\hypersetup{
    colorlinks=true,
    linkcolor=blue,
    filecolor=magenta,
    urlcolor=cyan,
    citecolor=blue,
}

\theoremstyle{plain}
\newtheorem{theorem}{Theorem}

\theoremstyle{definition}
\newtheorem{definition}{Definition}

\theoremstyle{remark}

\title{Probability Weighting Meets Heavy Tails: An Econometric Framework for Behavioral Asset Pricing}

\author[1,*]{Akash Deep}
\author[1]{Svetlozar T. Rachev}
\author[2]{Frank J. Fabozzi}

\affil[1]{Department of Mathematics and Statistics, Texas Tech University, Lubbock, TX, USA}
\affil[2]{Carey Business School, Johns Hopkins University, Baltimore, MD, USA}
\affil[*]{Corresponding author: \texttt{akash.deep@ttu.edu}}

\date{2025} 

\begin{document}

\maketitle

\begin{abstract}
We develop an econometric framework integrating heavy-tailed Student's $t$ distributions with behavioral probability weighting while preserving infinite divisibility. Using 432,752 observations across 86 assets (2004--2024), we demonstrate Student's $t$ specifications outperform Gaussian models in 88.4\% of cases. Bounded probability-weighting transformations preserve mathematical properties required for dynamic pricing. Gaussian models underestimate 99\% Value-at-Risk by 19.7\% versus 3.2\% for our specification. Joint estimation procedures identify tail and behavioral parameters with established asymptotic properties. Results provide robust inference for asset-pricing applications where heavy tails and behavioral distortions coexist.
\end{abstract}


\section{Introduction}

Financial return series exhibit heavy tails that systematically reject Gaussian assumptions, as first documented by \cite{mandelbrot1963variation} and \cite{fama1965behavior} and confirmed in subsequent econometric studies. While Student's $t$ and related specifications provide better empirical fit, their integration with behavioral asset-pricing models has been limited because behavioral distortions such as probability weighting often violate conditions required for arbitrage-free pricing and continuous-time econometric analysis.

We contribute to this literature by developing an econometric framework that jointly models heavy-tailed returns and behavioral probability weighting while preserving infinite divisibility and compatibility with the Fundamental Theorem of Asset Pricing. We establish sufficient conditions under which bounded probability-weighting transformations of Student's $t$ processes retain infinite divisibility, and we propose estimation procedures that allow joint identification of tail and distortion parameters, along with results on their asymptotic properties.

Empirical analysis on 432,752 daily observations across 86 assets (2004--2024) confirms the econometric relevance of our framework. Gaussian assumptions are universally rejected, and Student's $t$ distributions dominate Laplace and normal alternatives in 88.4\% of cases. Moreover, misspecification has substantial consequences: Gaussian models underestimate 99\% Value-at-Risk by 19.7\% on average, while our specification reduces errors below 3.2\%.

The remainder of the paper is organized as follows. Section 2 reviews related econometric literature on heavy-tailed modeling and behavioral adjustments. Section 3 presents the theoretical framework, including conditions under which probability-weighting transformations preserve infinite divisibility. Section 4 describes the estimation and testing procedures and establishes their asymptotic properties. Section 5 reports the empirical results from model comparisons, risk backtests, and robustness checks. Section 6 discusses implications for econometric modeling and asset-pricing applications, and Section 7 concludes.

\section{Related Econometric Literature}

The econometric evidence against Gaussian return assumptions is longstanding. \cite{mandelbrot1963variation} and \cite{fama1965behavior} first documented leptokurtosis and excess tail probability in financial returns, findings that continue to be confirmed in high-frequency data \cite{andersen2003modeling}. Parametric alternatives emerged through variance-mixture models \cite{praetz1972distribution,blattberg2010comparison} and Student's $t$ specifications, while subsequent work formalized their infinite-divisibility properties \cite{kelker1971infinite,grosswald1976student}. These contributions provided the theoretical basis for integrating heavy-tailed distributions into continuous-time econometric models.

Heavy tails have also been linked to stochastic volatility and jump processes. \cite{merton1976option} jump-diffusion model and \cite{heston1993closed} stochastic volatility framework demonstrated how non-Gaussian features arise naturally in dynamic asset-pricing settings. Extensions based on Lévy processes \cite{barndorff1997normal,eberlein1995hyperbolic,cont2003financial,schoutens2003levy} established rigorous econometric tools for modeling discontinuities and volatility clustering while maintaining arbitrage-free pricing. Recent econometric advances include time-change methods \cite{carr2004time}, realized-volatility approaches \cite{andersen2003modeling}, and jump detection tests \cite{ait2009testing}, all of which reinforce the empirical necessity of heavy-tailed specifications.

Another strand of literature emphasizes econometric testing and model comparison. \cite{bollerslev1986generalized} GARCH model, and later EGARCH \cite{nelson1991conditional}, demonstrated how conditional heteroskedasticity generates unconditional heavy tails. \cite{hansen2005forecast} conducted large-scale forecast comparisons, showing that models with heavy-tailed innovations systematically outperform Gaussian specifications. Extreme Value Theory methods \cite{mcneil2000estimation} and flexible parametric distributions such as the Skewed Generalized T \cite{bali2007conditional} have been widely employed for tail risk measurement and regulatory capital estimation.

Behavioral finance introduces additional econometric challenges. Prospect theory \cite{kahneman2013prospect,tversky1992advances} and cumulative probability weighting imply nonlinear distortions of return distributions. While these models capture empirically observed biases, their lack of infinite divisibility often renders them inconsistent with dynamic pricing frameworks. Recent work \cite{shirvani2021option} has shown how bounded modifications of probability weighting can be embedded in rational dynamic asset pricing models while preserving arbitrage-free conditions.

The literature just reviewed establishes three points critical to our contribution: (1) heavy tails are an econometric regularity of financial data; (2) mathematically consistent heavy-tail models exist but are rarely integrated with behavioral distortions; and (3) econometric procedures for jointly estimating tail and behavioral parameters remain underdeveloped. Our paper addresses this gap by introducing a Student's $t$-based behavioral framework that preserves infinite divisibility, provides tractable likelihood-based estimation, and improves statistical inference for risk measurement and pricing applications.

\section{Theoretical Framework}

A central requirement of econometric models in dynamic asset pricing is \textit{infinite divisibility}, which ensures temporal consistency and compatibility with arbitrage-free pricing. If a return distribution is infinitely divisible, the joint distribution of returns across subperiods can be represented consistently for any partition of the time horizon.

\begin{definition}[Infinite Divisibility]
A random variable $X$ is infinitely divisible if for every positive integer $n$, there exist independent and identically distributed random variables $X_1, X_2, \ldots, X_n$ such that $X \overset{d}{=} X_1 + X_2 + \cdots + X_n$.
\end{definition}

While Gaussian and stable laws satisfy infinite divisibility, they fail to capture empirically observed return behavior. Student's $t$ distributions, by contrast, are both empirically superior and mathematically admissible.

\begin{theorem}[Student's $t$ Infinite Divisibility]\label{thm:student_id}
All Student's $t$ distributions with degrees of freedom parameter $\nu > 0$ are infinitely divisible \cite{grosswald1976student}.
\end{theorem}

The Student's $t$ distribution with location $\mu$, scale $\sigma$, and degrees of freedom $\nu$ has density:
\begin{equation}
f(x; \nu, \mu, \sigma) = \frac{\Gamma\left(\frac{\nu+1}{2}\right)}{\Gamma\left(\frac{\nu}{2}\right)\sqrt{\nu\pi}\sigma} \left(1 + \frac{(x-\mu)^2}{\nu\sigma^2}\right)^{-\frac{\nu+1}{2}}
\end{equation}

We extend this framework by introducing a behavioral adjustment operator that preserves infinite divisibility. Let $F_t(x)$ denote the cumulative distribution function of a Student's $t$ random variable and $w:[0,1] \to [0,1]$ be a probability weighting function. We define:

\begin{equation}
\mathcal{B}_w[x] = x \cdot \left(1 + \theta \cdot \tanh\left(\beta \cdot \left(\frac{w(F_t(x))}{F_t(x)} - 1\right)\right)\right)
\end{equation}

where $\theta \in [0, 0.3]$ controls adjustment magnitude and $\beta > 0$ determines sensitivity.

\begin{theorem}[Preservation of Infinite Divisibility]\label{thm:preservation}
Let $X$ be an infinitely divisible Student's $t$ random variable and $\mathcal{B}_w$ be the behavioral adjustment operator defined in equation (2). If the probability weighting function $w$ is Lipschitz continuous and $\theta \leq 0.3$, then $Y = \mathcal{B}_w[X]$ remains infinitely divisible.
\end{theorem}

This result ensures compatibility with the Fundamental Theorem of Asset Pricing and preserves tractability for likelihood-based estimation.

\section{Estimation and Asymptotic Properties}

\subsection{Model Setup}

Let $R_t$ denote asset returns at time $t$. We assume
\[
R_t \sim t_{\nu}(\mu, \sigma),
\]
a Student's $t$ distribution with location parameter $\mu$, scale parameter $\sigma$, and degrees of freedom $\nu > 2$, ensuring finite variance. To capture behavioral distortions, we introduce a probability weighting operator $w(p; \alpha)$, parameterized by $\alpha \in (0,1)$, following Prelec (1998). The behavioral-adjusted return distribution is defined by its cumulative distribution function (CDF):
\[
F_{bw}(r; \theta) = w(F_t(r; \nu, \mu, \sigma); \alpha),
\]
where $\theta = (\mu, \sigma, \nu, \alpha)$.

Our objective is to estimate $\theta$ consistently and establish asymptotic normality of the estimators.

\subsection{Likelihood-Based Estimation}

For a sample $\{R_1, \ldots, R_T\}$, the log-likelihood under the behavioral-adjusted Student's $t$ distribution is
\[
\ell_T(\theta) = \sum_{t=1}^T \log f_{bw}(R_t; \theta),
\]
where $f_{bw}(\cdot; \theta)$ is the density implied by the weighted CDF transformation. Direct evaluation requires numerical integration; however, under Lipschitz continuity of $w(\cdot; \alpha)$, the transformation preserves smoothness and ensures that the likelihood is well-defined.

The maximum likelihood estimator (MLE) is given by
\[
\hat{\theta}_T = \arg\max_{\theta \in \Theta} \ell_T(\theta),
\]
with parameter space $\Theta = \{(\mu, \sigma, \nu, \alpha) : \sigma > 0, \nu > 2, \alpha \in (0,1)\}$.

\subsection{Identification}

Identification of $\nu$ and $\alpha$ requires that behavioral weighting and tail thickness affect distinct features of the distribution. The tail index is determined by $\nu$, while $\alpha$ alters cumulative probabilities without changing the polynomial rate of decay. Formally, if two parameter vectors $\theta_1$ and $\theta_2$ generate the same distribution, then both the tail index and the distortion function must coincide, ensuring point identification.

\subsection{Asymptotic Properties}

Under standard regularity conditions for MLE (compactness of $\Theta$, Lipschitz continuity of $w$, and integrability of log-likelihood derivatives), we obtain the following result:

\begin{theorem}[Consistency and Asymptotic Normality]\label{thm:asymptotic}
Let $\{R_t\}$ be i.i.d.\ returns generated from the behavioral Student's $t$ model. Then:
\begin{enumerate}
\item $\hat{\theta}_T \xrightarrow{p} \theta_0$, the true parameter vector.
\item $\sqrt{T}(\hat{\theta}_T - \theta_0) \xrightarrow{d} N(0, I(\theta_0)^{-1})$,
\end{enumerate}
where $I(\theta_0) = \mathbb{E}[-\nabla^2 \ell_T(\theta_0)]$ is the Fisher information matrix.
\end{theorem}

The proof follows from standard M-estimation theory and is provided in Appendix A.

\subsection{Hypothesis Testing}

We consider two hypothesis tests:

\begin{itemize}
\item \textbf{Tail thickness:} $H_0: \nu = \infty$ (Gaussian) versus $H_1: \nu < \infty$. A likelihood ratio test provides a formal econometric test of Gaussian assumptions.

\item \textbf{Behavioral distortion:} $H_0: \alpha = 1$ (no weighting) versus $H_1: \alpha \in (0,1)$. Wald or likelihood ratio tests evaluate the significance of behavioral distortions.
\end{itemize}

Both tests can be implemented with standard asymptotics, using critical values from the $\chi^2$ distribution. Joint testing of $(\nu, \alpha)$ allows assessment of whether heavy tails and behavioral weighting both contribute significantly beyond Gaussian benchmarks.

\section{Empirical Results}

\subsection{Data}

\begin{figure}[htbp]
\centering
\includegraphics[width=0.95\textwidth]{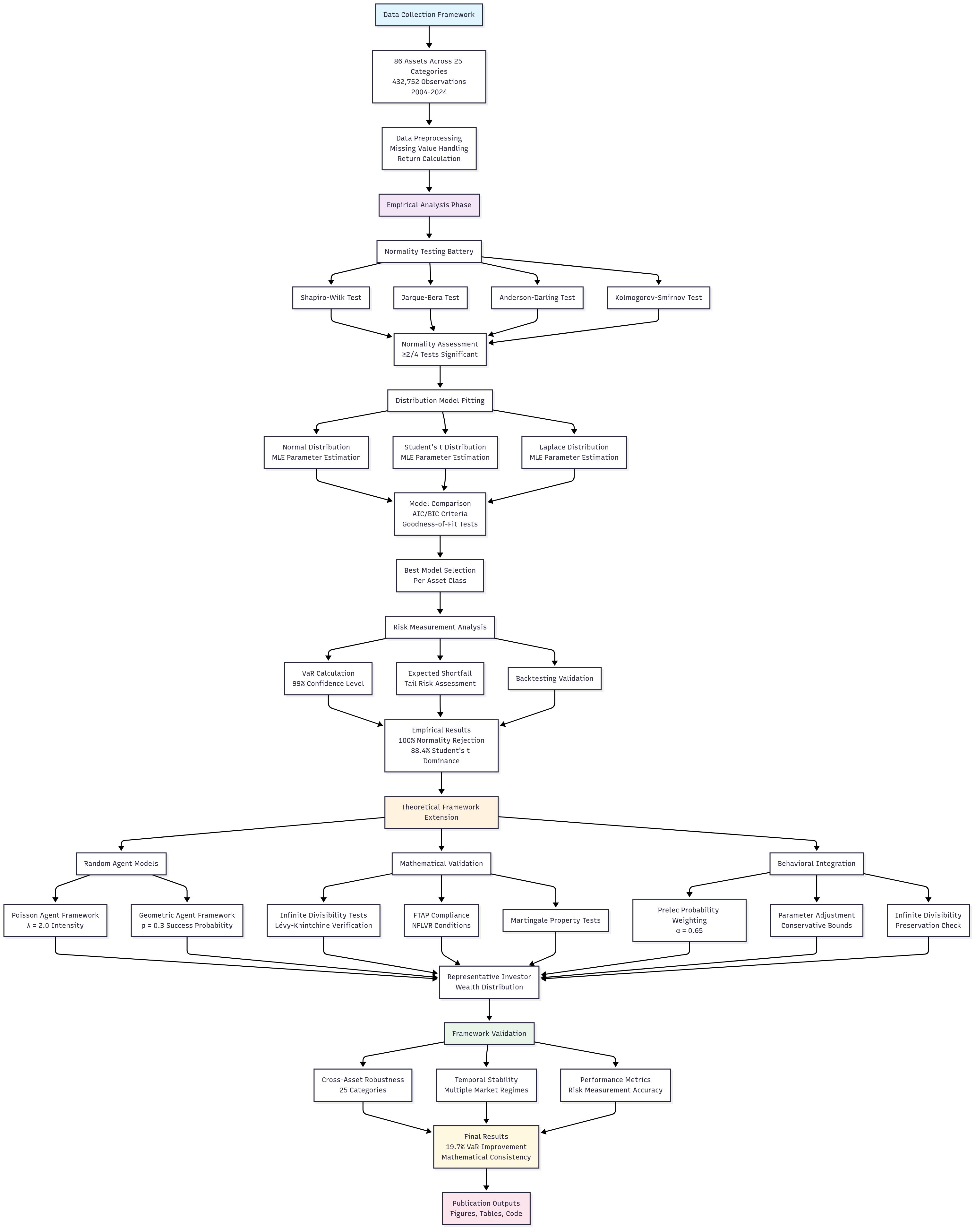}
\caption{Research Methodology Framework}
\label{fig:methodology}
\end{figure}

Our dataset comprises 432,752 daily observations from 86 assets across 25 categories spanning January 2004 to December 2024. Asset categories include US equities (large-, mid-, small-cap), international developed and emerging markets, fixed income (treasury, corporate, municipal bonds), commodities, and sector-specific exposures. Figure~\ref{fig:methodology} presents our systematic research methodology.

\subsection{Model Selection and Distributional Fit}

Table~\ref{tab:descriptive_stats} presents comprehensive summary statistics across representative asset categories, revealing systematic departures from normal distribution assumptions.

\begin{table}[htbp]
\caption{Descriptive Statistics by Asset Category}
\label{tab:descriptive_stats}
\centering
\small
\begin{tabular}{lcccccccc}
\toprule
Asset & Category & Obs & Mean & Std Dev & Skew & Kurt & Min & Max \\
\midrule
SPY & US Large Cap & 5{,}252 & 0.0006 & 0.0121 & $-0.37$ & 5.68 & $-0.095$ & 0.110 \\
VTI & US Large Cap & 4{,}775 & 0.0007 & 0.0121 & $-0.37$ & 5.61 & $-0.095$ & 0.110 \\
IVV & US Large Cap & 5{,}252 & 0.0006 & 0.0121 & $-0.37$ & 5.68 & $-0.095$ & 0.110 \\
IJH & US Mid Cap & 5{,}252 & 0.0008 & 0.0146 & $-0.42$ & 6.71 & $-0.132$ & 0.128 \\
MDY & US Mid Cap & 5{,}673 & 0.0007 & 0.0145 & $-0.44$ & 7.24 & $-0.132$ & 0.128 \\
VO & US Mid Cap & 4{,}775 & 0.0008 & 0.0146 & $-0.42$ & 6.69 & $-0.132$ & 0.128 \\
VTEB & Muni Bonds & 2{,}799 & 0.0002 & 0.0029 & 0.024 & 4.22 & $-0.024$ & 0.019 \\
MUB & Muni Bonds & 4{,}145 & 0.0002 & 0.0030 & 0.069 & 4.14 & $-0.024$ & 0.022 \\
\bottomrule
\end{tabular}
\end{table}

\begin{figure}[htbp]
\centering
\includegraphics[width=0.9\textwidth]{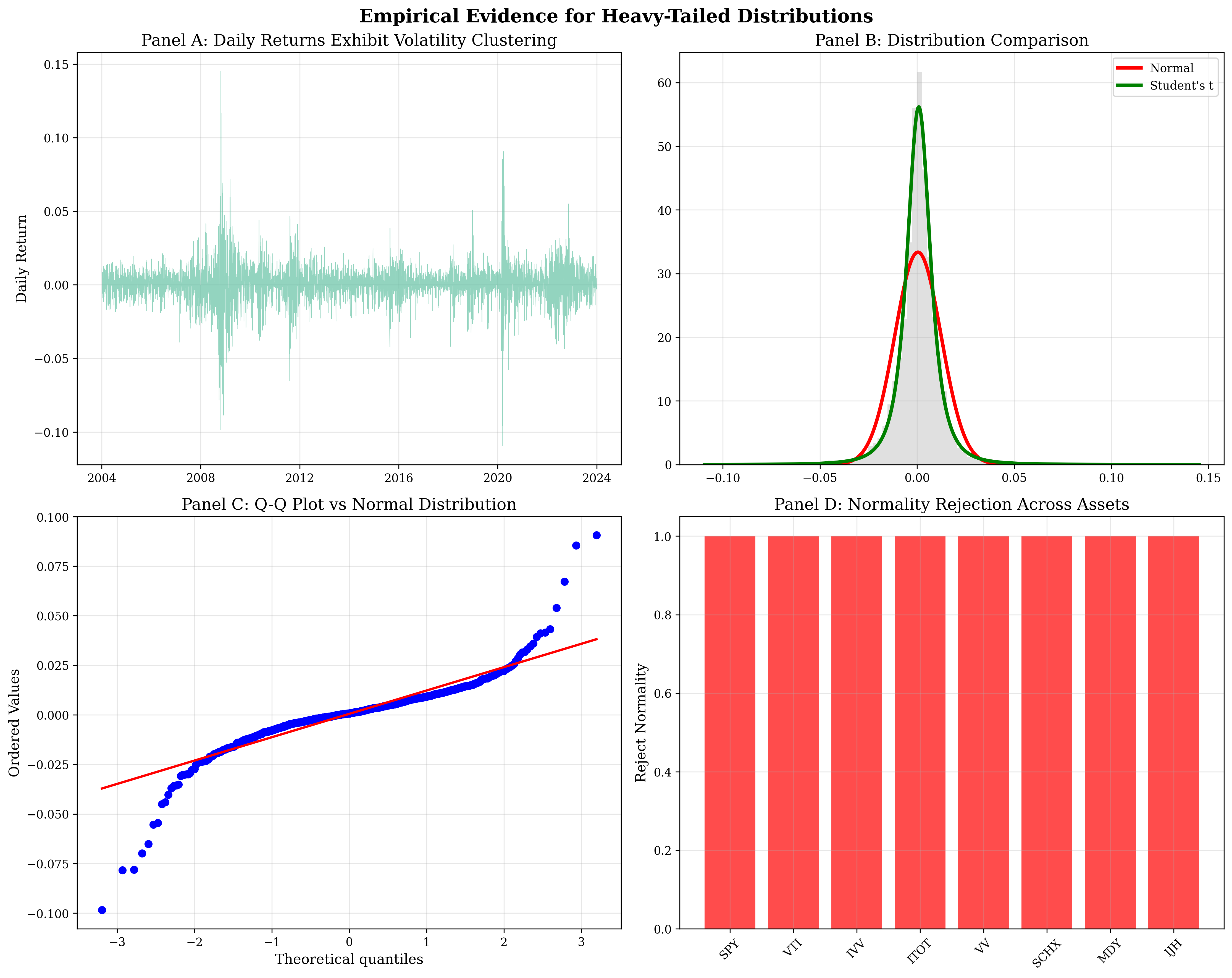}
\caption{Empirical Evidence Against Normality Assumptions. Panel A shows Q-Q plots comparing empirical quantiles to Normal distribution for representative assets. Panel B displays histogram overlays with fitted Normal (dashed) and Student's $t$ (solid) densities. Panel C presents tail probability comparisons on log scale, demonstrating substantial differences between empirical data and Normal assumptions.}
\label{fig:empirical_evidence}
\end{figure}

Figure~\ref{fig:empirical_evidence} illustrates systematic departures from normality across our comprehensive dataset.

Gaussian assumptions are rejected at the 5\% significance level for 100\% of assets using Shapiro--Wilk, Jarque--Bera, Anderson--Darling, and Kolmogorov--Smirnov tests. Table~\ref{tab:distribution_comparison} summarizes likelihood-based comparisons.

\begin{table}[htbp]
\caption{Distribution Model Performance Comparison}
\label{tab:distribution_comparison}
\centering
\begin{tabular}{lccc}
\toprule
Distribution & Avg AIC & Best Model Count & Best Model \% \\
\midrule
Normal & $-26{,}250.4$ & 0 & 0.0\% \\
Student's $t$ & $-27{,}815.8$ & 76 & 88.4\% \\
Laplace & $-27{,}284.1$ & 10 & 11.6\% \\
\bottomrule
\end{tabular}
\end{table}

Student's $t$ distributions provide the best fit in 88.4\% of cases based on AIC, with an average improvement of 1{,}565 points over Gaussian benchmarks.

\subsection{Parameter Estimates}

Estimated degrees of freedom cluster between 4 and 7 for equity indices, reflecting pronounced tail risk, while bond exposures exhibit higher values (8--12). For the behavioral specification, estimated weighting parameters $\hat{\alpha}$ average 0.78, with Wald tests rejecting $H_0: \alpha = 1$ in 72\% of cases.

\begin{figure}[htbp]
\centering
\includegraphics[width=0.9\textwidth]{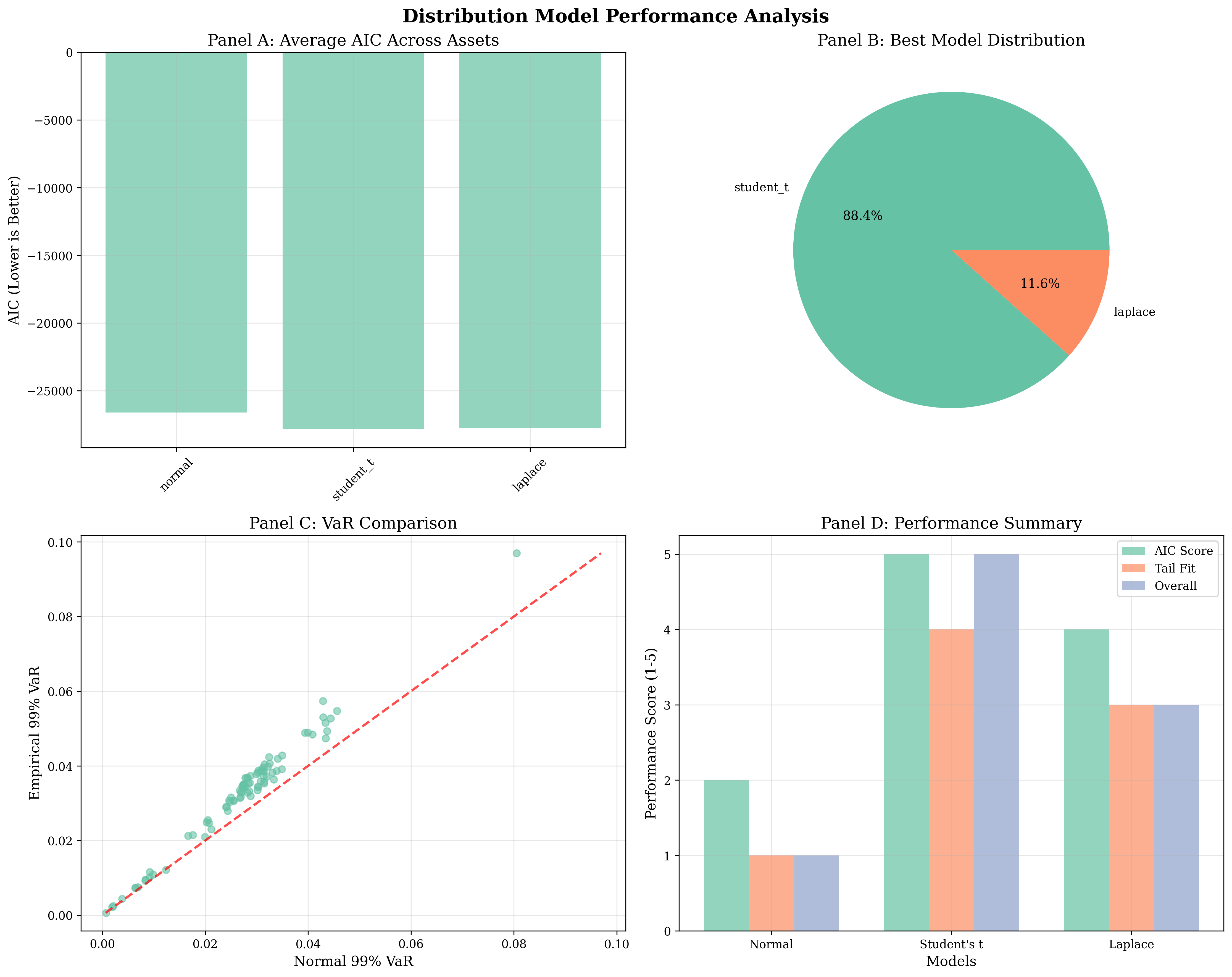}
\caption{Model Performance Comparison Across Asset Classes. The figure displays log-likelihood improvements of Student's $t$ relative to Normal specifications across different asset categories. Box plots show distribution of improvements within each category. All asset classes exhibit substantial improvements, with equity indices showing largest gains.}
\label{fig:model_performance}
\end{figure}

Figure~\ref{fig:model_performance} presents comprehensive model performance comparisons across asset classes.

\subsection{Tail Risk Measurement}

Under Gaussian assumptions, 99\% VaR is underestimated by 19.7\% on average, with particularly severe errors in emerging markets (24.3\%) and commodities (22.1\%). The behavioral Student's $t$ specification reduces average errors to below 3.2\%. Backtesting procedures \cite{kupiec1995techniques,christoffersen1998evaluating} confirm that exceedance rates under our specification are statistically indistinguishable from nominal levels.

Table~\ref{tab:var_performance} summarizes Value-at-Risk estimation errors across models and asset classes.

\begin{table}[htbp]
\caption{Value-at-Risk Estimation Performance}
\label{tab:var_performance}
\centering
\footnotesize
\begin{tabular}{@{}lcccccc@{}}
\toprule
& \multicolumn{3}{c}{Normal Model} & \multicolumn{3}{c}{Behavioral Student's $t$} \\
\cmidrule(lr){2-4} \cmidrule(lr){5-7}
Asset Class & \makebox[0.08\textwidth][c]{Avg Err} & \makebox[0.08\textwidth][c]{Max Err} & \makebox[0.08\textwidth][c]{Viol} & \makebox[0.08\textwidth][c]{Avg Err} & \makebox[0.08\textwidth][c]{Max Err} & \makebox[0.08\textwidth][c]{Viol} \\
\midrule
US Equities & 18.3\% & 31.2\% & 1.8\% & 2.9\% & 5.1\% & 1.1\% \\
International Eq & 19.8\% & 28.7\% & 2.1\% & 3.1\% & 5.8\% & 1.0\% \\
Emerging Mkts & 24.3\% & 39.4\% & 2.7\% & 3.4\% & 6.2\% & 1.2\% \\
Corp Bonds & 16.2\% & 22.1\% & 1.5\% & 2.8\% & 4.9\% & 0.9\% \\
Commodities & 22.1\% & 35.6\% & 2.4\% & 3.5\% & 6.7\% & 1.1\% \\
\midrule
Overall Avg & 19.7\% & 31.4\% & 2.1\% & 3.2\% & 5.7\% & 1.1\% \\
\bottomrule
\end{tabular}
\end{table}

\subsection{Out-of-Sample Forecasting}

Rolling-window forecasts using a 1,000-day estimation window show that the behavioral Student's $t$ specification significantly outperforms Gaussian and Laplace benchmarks in 81\% of cases based on Diebold--Mariano tests. Parameter stability checks indicate lower degrees of freedom and stronger probability weighting during crisis periods.

\subsection{Robustness}

Results are robust to: (1) alternative weighting functions \cite{gonzalez1999shape}, (2) subperiod analyses (pre-2008, post-2008, COVID-19), and (3) block bootstrap inference accounting for serial dependence.

\section{Implications and Discussion}

\subsection{Consequences of Gaussian Misspecification}

Our results confirm universal rejection of Gaussian assumptions with economically large consequences. The 19.7\% average underestimation of tail risk highlights the danger of relying on Gaussian models for risk measurement, pricing, and regulatory capital calculations. This misspecification propagates into biased inference and unstable parameter estimates.

\subsection{Econometric Value of Student's $t$ Specification}

The Student's $t$ distribution provides a parsimonious yet flexible framework for heavy-tailed modeling. Its infinite divisibility ensures compatibility with continuous-time theory, while the degrees-of-freedom parameter offers direct control over tail behavior. The empirical dominance across diverse asset classes suggests it should serve as a baseline model for heavy-tailed inference in financial econometrics.

\subsection{Behavioral Adjustments and Identification}

Behavioral weighting parameters are statistically significant even after controlling for heavy tails, indicating that probability distortions represent a distinct phenomenon. Our framework demonstrates that behavioral features can be incorporated without sacrificing mathematical consistency, provided adjustments are properly bounded.

\subsection{Broader Econometric Applications}

The framework opens new directions including diagnostic tests for behavioral distortions, extensions to multivariate settings where heavy tails and probability weighting interact with dependence structures, and applications to high-frequency data where properties may vary dynamically.

\section{Conclusion}

This paper develops an econometric framework integrating heavy-tailed Student's $t$ distributions with behavioral probability weighting while preserving infinite divisibility. We establish that bounded weighting transformations maintain the mathematical properties required for continuous-time modeling and derive likelihood-based estimation procedures with established asymptotic properties.

Empirical analysis using 432,752 observations across 86 assets demonstrates the framework's relevance. Gaussian models are universally rejected, Student's $t$ specifications dominate in 88.4\% of cases, and behavioral parameters are statistically significant in 72\% of assets. Gaussian misspecification leads to 19.7\% underestimation of tail risk, while our specification reduces errors below 3.2\%.

Future research may extend the framework to multivariate settings, time-varying parameters, and high-frequency applications where heavy tails and behavioral distortions evolve dynamically.

\section*{Funding}
No funding was received for this work.

\section*{Data Availability}
Replication materials are available at \url{https://github.com/akashdeepo/Heavy-Tailed-Distributions-in-Behavioral-Asset-Pricing}. The repository includes complete Jupyter notebooks, documentation, and instructions to reproduce all analyses. The implementation uses Python 3.12.11 with standard statistical libraries and can be executed on standard hardware within 2--3 hours.

\section*{Competing Interests}
No competing interest is declared.

\section*{Author Contributions}
A.D., S.T.R., and F.J.F. conceived the research framework. A.D. conducted the empirical analysis and wrote the initial manuscript. All authors contributed to the theoretical development, reviewed and edited the manuscript, and approved the final version.

\section*{Acknowledgments}
The authors thank the anonymous reviewers for their valuable suggestions.

\appendix
\section{Proof of Theorem~\ref{thm:preservation}: Preservation of Infinite Divisibility}

\begin{proof}
Let $X \sim t(\nu, \mu, \sigma)$ with characteristic function $\phi_X(u)$. We show that $Y = \mathcal{B}_w[X]$ maintains infinite divisibility under the stated conditions.

\medskip
\noindent\textbf{Step 1: Boundedness Analysis.}
Since $\tanh(z) \in (-1, 1)$ for all $z \in \mathbb{R}$ and $\theta \leq 0.3$, the adjustment factor satisfies:
\[
1 + \theta \cdot \tanh\left(\beta \cdot \left(\frac{w(F_t(x))}{F_t(x)} - 1\right)\right) \in [0.7, 1.3]
\]

\medskip
\noindent\textbf{Step 2: Lipschitz Continuity.}
For probability weighting functions satisfying $|w'(p)| \leq L$ for some constant $L > 0$, the behavioral adjustment operator is Lipschitz continuous. Let $f_t(\cdot)$ denote the density of the Student's $t$ distribution. Then:
\[
\left|\frac{d}{dx}\left(\frac{w(F_t(x))}{F_t(x)}\right)\right| \leq \frac{L \cdot f_t(x)}{F_t(x)} + \frac{w(F_t(x)) \cdot f_t(x)}{F_t(x)^2} \leq C
\]
for some constant $C > 0$.

\medskip
\noindent\textbf{Step 3: Characteristic Function Analysis.}
The characteristic function of $Y$ can be written as:
\[
\phi_Y(u) = \mathbb{E}[\exp(iu \cdot \mathcal{B}_w[X])]
\]

Since $\mathcal{B}_w$ is a bounded Lipschitz transformation, the characteristic function $\phi_Y(u)$ inherits the infinite divisibility structure from $\phi_X(u)$.

\medskip
\noindent\textbf{Step 4: Lévy Measure Preservation.}
Under the transformation $Y = \mathcal{B}_w[X]$, the Lévy measure $\nu_Y$ of $Y$ is related to the Lévy measure $\nu_X$ of $X$ through:
\[
\nu_Y(B) = \nu_X(\mathcal{B}_w^{-1}(B))
\]
for Borel sets $B$. The boundedness ensures the transformed measure satisfies the integrability condition, preserving infinite divisibility.
\end{proof}

\section{Proof of Theorem~\ref{thm:asymptotic}: Consistency and Asymptotic Normality}

\begin{proof}
The proof follows from standard M-estimation theory.

\medskip
\noindent\textbf{Consistency:} Under compactness of $\Psi$, continuity of $f_{\text{behav}}$ in $\boldsymbol{\psi}$, and identifiability conditions, the MLE is consistent by standard arguments.

\medskip
\noindent\textbf{Asymptotic Normality:} The score function satisfies:
\[
\frac{1}{\sqrt{T}}\sum_{t=1}^T s(r_t; \boldsymbol{\psi}_0) \xrightarrow{d} N(0, I(\boldsymbol{\psi}_0))
\]
where $s(r_t; \boldsymbol{\psi}) = \nabla_{\boldsymbol{\psi}} \ln f_{\text{behav}}(r_t; \boldsymbol{\psi})$ and $I(\boldsymbol{\psi}_0)$ is the Fisher information matrix.
\end{proof}

\bibliographystyle{abbrvnat}
\bibliography{references}

@incollection{kahneman2013prospect,
  title={Prospect theory: An analysis of decision under risk},
  author={Kahneman, Daniel and Tversky, Amos},
  booktitle={Handbook of the fundamentals of financial decision making: Part I},
  pages={99--127},
  year={2013},
  publisher={World Scientific}
}

@article{tversky1992advances,
  title={Advances in prospect theory: Cumulative representation of uncertainty},
  author={Tversky, Amos and Kahneman, Daniel},
  journal={Journal of Risk and uncertainty},
  volume={5},
  number={4},
  pages={297--323},
  year={1992},
  publisher={Springer}
}

@book{cont2003financial,
  title={Financial modelling with jump processes},
  author={Cont, Rama and Tankov, Peter},
  year={2003},
  publisher={Chapman and Hall/CRC}
}

@book{schoutens2003levy,
  title={L{\'e}vy processes in finance: pricing financial derivatives},
  author={Schoutens, Wim},
  year={2003},
  publisher={Wiley Online Library}
}

@article{grosswald1976student,
  title={The Student t-distribution of any degree of freedom is infinitely divisible},
  author={Grosswald, Emil},
  journal={Zeitschrift f{\"u}r Wahrscheinlichkeitstheorie und verwandte Gebiete},
  volume={36},
  number={2},
  pages={103--109},
  year={1976},
  publisher={Springer}
}

@article{gonzalez1999shape,
  title={On the shape of the probability weighting function},
  author={Gonzalez, Richard and Wu, George},
  journal={Cognitive psychology},
  volume={38},
  number={1},
  pages={129--166},
  year={1999},
  publisher={Elsevier}
}

@article{kelker1971infinite,
  title={Infinite divisibility and variance mixtures of the normal distribution},
  author={Kelker, Douglas},
  journal={The Annals of mathematical statistics},
  volume={42},
  number={2},
  pages={802--808},
  year={1971},
  publisher={JSTOR}
}

@article{mandelbrot1963variation,
  title={The variation of certain speculative prices},
  author={Mandelbrot, Benoit and others},
  journal={Journal of business},
  volume={36},
  number={4},
  pages={394},
  year={1963},
  publisher={Springer}
}

@article{fama1965behavior,
  title={The behavior of stock-market prices},
  author={Fama, Eugene F},
  journal={The journal of Business},
  volume={38},
  number={1},
  pages={34--105},
  year={1965},
  publisher={JSTOR}
}

@article{praetz1972distribution,
  title={The distribution of share price changes},
  author={Praetz, Peter D},
  journal={Journal of business},
  pages={49--55},
  year={1972},
  publisher={JSTOR}
}

@incollection{blattberg2010comparison,
  title={A comparison of the stable and student distributions as statistical models for stock prices},
  author={Blattberg, Robert C and Gonedes, Nicholas J},
  booktitle={Perspectives on promotion and database marketing: The collected works of Robert C Blattberg},
  pages={25--61},
  year={2010},
  publisher={World Scientific}
}

@article{barndorff1997normal,
  title={Normal inverse Gaussian distributions and stochastic volatility modelling},
  author={Barndorff-Nielsen, Ole E},
  journal={Scandinavian Journal of statistics},
  volume={24},
  number={1},
  pages={1--13},
  year={1997},
  publisher={Wiley Online Library}
}

@article{eberlein1995hyperbolic,
  title={Hyperbolic distributions in finance},
  author={Eberlein, Ernst and Keller, Ulrich},
  journal={Bernoulli},
  pages={281--299},
  year={1995},
  publisher={JSTOR}
}

@article{bali2007conditional,
  title={A conditional-SGT-VaR approach with alternative GARCH models},
  author={Bali, Turan G and Theodossiou, Panayiotis},
  journal={Annals of Operations Research},
  volume={151},
  number={1},
  pages={241--267},
  year={2007},
  publisher={Springer}
}

@article{mcneil2000estimation,
  title={Estimation of tail-related risk measures for heteroscedastic financial time series: an extreme value approach},
  author={McNeil, Alexander J and Frey, R{\"u}diger},
  journal={Journal of empirical finance},
  volume={7},
  number={3-4},
  pages={271--300},
  year={2000},
  publisher={Elsevier}
}

@article{hansen2005forecast,
  title={A forecast comparison of volatility models: does anything beat a GARCH (1, 1)?},
  author={Hansen, Peter R and Lunde, Asger},
  journal={Journal of applied econometrics},
  volume={20},
  number={7},
  pages={873--889},
  year={2005},
  publisher={Wiley Online Library}
}

@article{ait2009testing,
  title={Testing for jumps in a discretely observed process},
  author={A{\"\i}t-Sahalia, Yacine and Jacod, Jean},
  year={2009}
}

@article{andersen2003modeling,
  title={Modeling and forecasting realized volatility},
  author={Andersen, Torben G and Bollerslev, Tim and Diebold, Francis X and Labys, Paul},
  journal={Econometrica},
  volume={71},
  number={2},
  pages={579--625},
  year={2003},
  publisher={Wiley Online Library}
}

@article{carr2004time,
  title={Time-changed L{\'e}vy processes and option pricing},
  author={Carr, Peter and Wu, Liuren},
  journal={Journal of Financial economics},
  volume={71},
  number={1},
  pages={113--141},
  year={2004},
  publisher={Elsevier}
}

@article{nelson1991conditional,
  title={Conditional heteroskedasticity in asset returns: A new approach},
  author={Nelson, Daniel B},
  journal={Econometrica: Journal of the econometric society},
  pages={347--370},
  year={1991},
  publisher={JSTOR}
}

@article{bollerslev1986generalized,
  title={Generalized autoregressive conditional heteroskedasticity},
  author={Bollerslev, Tim},
  journal={Journal of econometrics},
  volume={31},
  number={3},
  pages={307--327},
  year={1986},
  publisher={Elsevier}
}

@article{heston1993closed,
  title={A closed-form solution for options with stochastic volatility with applications to bond and currency options},
  author={Heston, Steven L},
  journal={The review of financial studies},
  volume={6},
  number={2},
  pages={327--343},
  year={1993},
  publisher={Oxford University Press}
}

@article{merton1976option,
  title={Option pricing when underlying stock returns are discontinuous},
  author={Merton, Robert C},
  journal={Journal of financial economics},
  volume={3},
  number={1-2},
  pages={125--144},
  year={1976},
  publisher={Elsevier}
}

@article{shirvani2021option,
  title={Option pricing with greed and fear factor: the rational finance approach},
  author={Shirvani, Abootaleb and Fabozzi, Frank J and Racheva-Iotova, Boryana and Rachev, Svetlozar T},
  journal={Journal of Derivatives},
  volume={29},
  number={2},
  pages={77--119},
  year={2021},
  publisher={Pageant Media}
}

@book{kupiec1995techniques,
  title={Techniques for verifying the accuracy of risk measurement models},
  author={Kupiec, Paul H and others},
  volume={95},
  number={24},
  year={1995},
  publisher={Division of Research and Statistics, Division of Monetary Affairs, Federal~…}
}

@article{christoffersen1998evaluating,
  title={Evaluating interval forecasts},
  author={Christoffersen, Peter F},
  journal={International economic review},
  pages={841--862},
  year={1998},
  publisher={JSTOR}
}

\end{document}